\documentclass[useAMS,usenatbib,preprint]{mn2e}
\usepackage{graphicx}
\usepackage[dvips]{color}
\usepackage{amsmath}
\usepackage{amssymb}
\usepackage{ulem}
\usepackage{multicol}
\title[Envelope Pollution of Gas Giants by Icy Planetesimals]{Gas Giant Formation with Small Cores Triggered by Envelope Pollution by Icy Planetesimals}
\author[Y. Hori and M. Ikoma]{Y.Hori$^{1}$\thanks{E-mail:yasunori.hori@nao.ac.jp}; M. Ikoma$^{1}$
\\$^{1}$
Department of Earth and Planetary Sciences, Tokyo Institute of Technology, 2-12-1 Ookayama, Meguro-ku, Tokyo 152-8551, Japan}
\begin{document}

\date{Accepted 2011 May 26. Received 2011 April 20}

\pagerange{\pageref{firstpage}--\pageref{lastpage}} \pubyear{2011}

\maketitle

\label{firstpage}

\begin{abstract}
We have investigated how envelope pollution by icy planetesimals affects the critical core mass for gas giant formation and the gas accretion time-scales. In the core-accretion model, runaway gas accretion is triggered after a core reaches a critical core mass. All the previous studies on the core-accretion model assumed that the envelope has the solar composition uniformly. In fact, the envelope is likely polluted by evaporated materials of icy planetesimals because icy planetesimals going through the envelope experience mass loss via strong ablation and most of their masses are deposited in the deep envelope. In this paper, we have demonstrated that envelope pollution in general lowers the critical core masses and hastens gas accretion on to the protoplanet because of the increase in the molecular weight and reduction of adiabatic temperature gradient. Widely- and highly-polluted envelopes allow smaller cores to form massive envelopes before disc dissipation. Our results suggest that envelope pollution in the course of planetary accretion has the potential to trigger gas giant formation with small cores. We propose that it is necessary to take into account envelope pollution by icy planetesimals when we discuss gas giant formation based on the core-accretion model.
\end{abstract}

\begin{keywords}
accretion, accretion discs -- planets and satellites: formation.
\end{keywords}

\section{Introduction}

More than 540 exoplanets have been discovered so far\footnote{http://www.exoplanet.eu}. Dedicated planet surveys have revealed the diversity of giant planets outside the Solar System. Many Sun-like stars are known to harbor close-in giant planets called hot-Jupiters, in contrast to our Sun. Recently, direct imaging has shed light on the existence of distant extrasolar giant planets such as HR8799b, c, d, e \citep{Marois+08,Marois+10}, Fomalhaut b \citep{Kalas+08}, and Beta Pic b \citep{Lagrange+09}. A large number of exoplanets enable us to consider planet formation around various types of star through a statistical approach, which is often called population synthesis. In the population synthesis, planetary systems are built by putting together pieces of formation processes (e.g., solid and gas accretion, planetary migration and so on) in a Monte-Carlo way. Nowadays several groups have worked on the population synthesis independently \citep[e.g.][]{Ida+08,Alibert+11,Kennedy+08,Miguel+11}. Those population-synthesis studies succeed in reproducing the planetary mass-period distribution of detected extrasolar giant planets. This indicates that our basic picture of planetary formation is supported by observed exoplanets. On the other hand, it is also true that different population syntheses based on different assumptions as to planetary accretion and disc properties reproduce the same observed distribution. This highlights the need to understand underlying formation processes of giant planets more exactly. 

The core-accretion model is one of the promising models for giant planet formation \citep[e.g.][]{Mizuno80,Bodenheimer+86,Pollack+96}. In this model, once a core reaches a critical core mass through planetesimal accretion, runaway gas accretion is triggered and the envelope mass increases rapidly. The typical value of the critical core mass is thought of as being $\gtrsim 10 M_\oplus$ \citep{Pollack+96,Fortier+07,Fortier+09} to form a massive envelope within the disc lifetime suggested by observations \citep{Haisch+01,Hernandez+07}. This is, however, incompatible with the inferred mass of Jupiter's core of $\lesssim 10 M_\oplus$ \citep{Saumon+04}. The critical core mass depends on planetesimal accretion rate and opacities in the envelope \citep{Mizuno80,Ikoma+00,Rafikov06}. Fast accretion of planetesimals leads to form gas giants with large cores. Although slow rates of planetesimal accretion lead to smaller critical cores of $< 10 M_\oplus$, gas accretion on to small cores tends to be long. Reduction of grain opacities in the envelope can not only lower critical core masses but also hasten gas accretion on to the protoplanet \citep{Ikoma+00,Hubickyj+05,Papaloizou+05,Hori+10}. For example, \citet{Hori+10} showed that core masses needed to form massive envelopes before disc dissipation are as small as $1 M_\oplus$ in the extreme case of grain-free envelopes.
In fact, recent calculations on the dynamical behaviour of dust grains in the accreting envelope demonstrate that grain opacities in the envelope can be reduced
to be on average 1\% or less of those in the protoplanetary disc
\citep{Podolak03,Movshovitz+08,Movshovitz+10}.
The reason is that small grains initially suspended in the outer envelope quickly
grow large in size and then settle down into the deep envelope
where temperature is high enough so that grains evaporate.
Their results support that reduction of grain opacities can be a feasible idea
that relaxes the problem of the formation of gas giants with small cores.

Another factor that affects the critical core mass is considered in this paper. All the previous studies assumed that the composition of gas is solar throughout the envelope. In fact, the envelope is likely to be polluted by evaporated materials of icy planetesimals. Once a protoplanet grows up to be Mars-size, it becomes difficult for the protoplanet to capture planetesimals because the protoplanet enhances their random velocities through gravitational scatterings by itself. On the other hand, disruptive collisions between planetesimals with high relative velocities take place frequently and yield a large number of 10 to 100 m-sized fragments unless the planetesimals are larger than 100~km \citep{Inaba+03b,Kobayashi+10}. The protoplanet can capture such small-sized fragments because of
the enhanced collisional cross-section between the envelope and those fragments due to gas drag \citep{Inaba+03a,Benvenuto+08}. The fragments going through the envelope experience both melting and evaporation and lose most of their masses in the deep envelope before reaching the core \citep{Pollack+86,Podolak+88}.
As a result, envelope pollution by evaporated materials of icy planetesimals
occurs inevitably in the course of planetary accretion.
In addition, evaporation (or erosion) of the core itself is also expected to occur, 
as pointed out by \citet{Lissauer+95}.
This fact motivates us to consider giant planet formation taking into account
the effects of the heavy-element enrichment in the envelope.

Approximate solutions for the critical core mass analytically derived by previous studies tell that envelope pollution has significant impacts on the critical core mass: 
\begin{itemize}
\item[(a)] Wholly-radiative envelopes \citep{Stevenson82}:
	\begin{equation}
		M_\mathrm{crit} \propto \mu^{-\frac{12}{7}} \kappa^{\frac{3}{7}}
			\dot{M}^{\frac{3}{7}}_\mathrm{c} (\ln R_\mathrm{out})^{-\frac{3}{7}},
		\label{Mcrit_rad}
	\end{equation}
\item[(b)] Wholly convective envelopes \citep{Wuchterl93}:
	\begin{equation}
		M_\mathrm{crit} \propto \mu^{-\frac{3}{2}}~\Gamma^{\frac{3}{2}}_1
			\frac{\sqrt{\Gamma_1 - 4/3}}{(\Gamma_1 - 1)^2}~
			\rho_\mathrm{out}^{-\frac{1}{2}}~T_\mathrm{out}^{\frac{3}{2}},
	\label{Mcrit_conv}
	\end{equation}
\end{itemize}
where $M_\mathrm{crit}$ is the critical core mass,
$\mu$ the molecular weight, $\kappa$ the Rosseland mean opacity,
$\dot{M}_\mathrm{c}$ the planetesimal accretion rate,
and $\Gamma_1$ the adiabatic exponent.
The subscript "out" means the value at the outer edge of the envelope.
The envelope pollution causes increase in $\mu$ and reduction of $\nabla_\mathrm{ad}$ (or reduction of $\Gamma_1$). The reduction of $\nabla_\mathrm{ad}$ comes from dissociations of several kinds of molecules.
The two formulae indicate that both effects reduce the critical core mass. This is because those effects lead to reduction of the local pressure gradient that supports the gravity. On the other hand, the envelope pollution also causes rise in gas opacities as a negative effect.
The rise in gas opacities, mainly $\mathrm{H}_2\mathrm{O}$ that
absorbs infrared radiation, increases the critical core mass by increasing the temperature gradient that results in an increase in the pressure gradient. 

In this study, we show that envelope pollution by icy planetesimals lowers critical core masses in most cases and hastens gas accretion on to the protoplanet
despite the rise in gas opacities.
We present numerical procedures in Section 2, including our modelling for
envelope pollution by icy planetesimals.
Results of critical core masses and gas accretion
time-scales are shown in Section 3 and Section 4, respectively.
Discussions and conclusion are in the last two sections.

\section{Numerical Procedure \label{sec.2}}
	\subsection{Structure and Evolution of the Envelope \label{sec.2-1}}
\vspace{0.2cm}

We suppose a spherically-symmetric protoplanet that
consists of a gaseous envelope and a rigid core
with a constant density of $3.2~\mathrm{g}~\mathrm{cm}^{-3}$. 
The envelope is assumed to be in hydrostatic equilibrium.
Its structure is simulated with a set of the following basic equations
\citep[e.g.][]{Kippenhahn+94} that includes
the equation of hydrostatic equilibrium,
	\begin{equation}
		\frac{\partial P}{\partial M_r} = -\frac{GM_r}{4\pi r^4},
	\label{HE}
	\end{equation}
the equation of mass conservation,
	\begin{equation}
		\frac{\partial r}{\partial M_r} = \frac{1}{4\pi r^2 \rho},
	\label{mass_conv}
	\end{equation}
the equation of heat transfer,
	\begin{equation}
		\frac{\partial T}{\partial M_r} = \frac{T}{P}\frac{\partial P}{\partial M_r}\nabla,
	\label{heat_transfer}
	\end{equation}
and the equation of energy conservation,
	\begin{equation}
		\frac{\partial L_r}{\partial M_r} = \epsilon_\mathrm{acc} - T\frac{dS}{dt},
	\label{energy_conv}
	\end{equation}
where $M_r$ is the mass interior to radius $r$, $P$ the pressure, $T$ the temperature,
$\rho$ the density, $S$ the specific entropy, $L_r$ the energy flux crossing at a sphere of
radius $r$, $\nabla = d\log T/d\log P$, $G$ the gravitational constant,
$\epsilon_\mathrm{acc}$ the specific energy generation by planetesimal accretion,
and $t$ the time.

The dominant mechanism of heat transfer is chosen
by the Schwartzschild criterion for convective instability:

	\begin{equation}
		\left( \frac{\partial \ln T}{\partial \ln P} \right)_s 
		\equiv \nabla_\mathrm{ad} < \nabla_\mathrm{rad} ,
	\label{criterion}
	\end{equation}
where $\nabla_\mathrm{ad}$ is the adiabatic temperature gradient and
$\nabla_\mathrm{rad}$ the radiative temperature gradient given by
	\begin{equation}
 		\nabla_\mathrm{rad} = \frac{3}{16\pi ac G}\frac{\kappa P L_r}{M_r T^4},
	\end{equation}
where $c$ is the velocity of light, $a$ the radiation density constant, and
$\kappa$ the Rosseland mean opacity.
Opacities both of dust grains and gas and 
the equation of state for gaseous components that we use are mentioned
in Section~\ref{sec.2-2}.

The boundary conditions are given at
the core surface and the outer edge of the envelope as follows:

	\begin{equation}
		 r =  \left( \frac{3M_\mathrm{core}}{4\pi \rho_\mathrm{core}} \right)^{1/3}
		 ~\mathrm{and}~
		 L = L_\mathrm{core}
		 ~~\mathrm{at}~M_r = M_\mathrm{core},
	\label{inner_bc}
	\end{equation}
and 
	\begin{equation}
		 T = T_\mathrm{disc}~~\mathrm{and}~~\rho = \rho_\mathrm{disc}
		  ~~\mathrm{at}~M_r = M_\mathrm{p},
	\end{equation}
where $M_\mathrm{core}$ and $\rho_\mathrm{core}$ are the core mass and density,
respectively, $L_\mathrm{core}$ the luminosity at the core surface, $M_\mathrm{p}$ the protoplanet's total mass, and $\rho_\mathrm{disc}$ and $T_\mathrm{disc}$ are the density and temperature of the disc gas, respectively. The total mass of the planet, $M_\mathrm{p}$, is the mass of gas contained inside the outer edge of the envelope, which is defined by the smaller of the Bondi radius ($R_\mathrm{B}$)
and the Hill radius ($R_\mathrm{Hill}$):
	\begin{equation}
		R_\mathrm{B} = \frac{GM_\mathrm{p}}{c^2_s},~~
		R_\mathrm{Hill} = a_\mathrm{p}
			\left[ \frac{M_\mathrm{p}}{3(M_\mathrm{p}+M_*)} \right]^{1/3},
	\end{equation}
where $c_s$ is the sound velocity, $a_\mathrm{p}$ the semimajor axis of the protoplanet, and $M_*$ the mass of the central star.
In this paper, we use $a_\mathrm{p} = 5.2 \mathrm{AU},~
M_* = 1M_\odot,~T_\mathrm{disc} = 150\mathrm{K}~
\mathrm{,and}~\rho_\mathrm{disc} = 5.0 \times 10^{-11}~\mathrm{g}~\mathrm{cm}^{-3}$, unless otherwise noted.

We assume that the energy release by planetesimals
occurs in a narrow region on top of the core (i.e., $\epsilon_\mathrm{acc} = 0$).
This assumption seems to be incompatible with the situation
that planetesimals evaporate and deposit their kinetic energy before reaching the core.
However, such simplification affects the results little, because
the energy deposition of planetesimals occurs mostly in the deep, convective envelope, the structure of which is insensitive to where energy deposition occurs inside itself. On the other hand, we neglect the effect of sinking of dissolved materials of icy planetesimals in this study. Gravitational potential energy released by sinking of ablated materials makes a contribution to slow down gas accretion on to a proto-gas giant as pointed out by \citet{Pollack+96}. Whether the energy release by sinking of ablated materials occurs effectively in a wide region where planetesimals slow down via ablation is important for formation time-scales of gas giants. It will be considered in future calculations based on trajectories of planetesimals through a protoplanet's envelope.

\subsection{Envelope Pollution by Icy Planetesimals \label{sec.2-2}}
\vspace{0.2cm}

\begin{figure}
\includegraphics[width=0.9\hsize,clip]{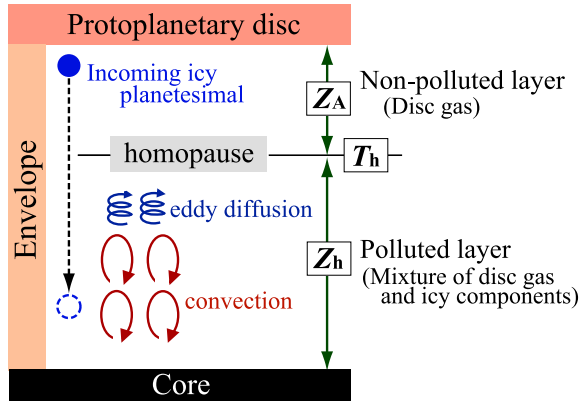}
\caption{Schematic picture of the envelope polluted by icy planetesimals.
The envelope has two-layer structure:
the non-polluted ($Z_\mathrm{A}$) and the polluted ($Z_\mathrm{h}$) layers,
where $Z_\mathrm{A}$ and $Z_\mathrm{h}$ are the mass fractions of
heavy elements for the disc gas and the mixture of the disc gas and icy planetesimals,
respectively. The boundary between the upper and the lower layers is defined by
the homopause temperature ($T_\mathrm{h}$).
\label{fig1}}
\end{figure}

In this study, we consider that the envelope consists of
two components, materials of disc gas (component A) and those of icy planetesimals (component B). The chemical compositions of component A and B
are assumed to be solar and comet-Halley-like, respectively.
The mass fractions of hydrogen, helium, and the others in the solar abundances
are $X_\mathrm{A} = 0.711,~Y_\mathrm{A} = 0.274,~\mathrm{and}~Z_\mathrm{A} = 0.015$,
respectively \citep{Lodders+09}, while those in comet Halley are
$X_\mathrm{B} = 0.06,~Y_\mathrm{B} = 0.00,~\mathrm{and}~Z_\mathrm{B} = 0.94$
\citep{Mumma+93}. 

We consider that the envelope has two-layer structure
(see Figure \ref{fig1}).
The upper layer consists only of component A because disc gas flows into the outer part of the envelope. The lower layer consists of component B in addition to A. The reason is as follows. While going through the envelope, planetesimals experience mass loss via
strong ablation. Most of their masses are deposited in the deep envelope \citep{Pollack+86,Podolak+88}.
In the lower envelope, convection occurs because of molecular dissociation
and high gas opacity due to the bound-free absorption of hydrogen.
The evaporated materials are thus stirred effectively by convection, so that
the chemical composition becomes uniform in the lower layer of the envelope.
Such a boundary between convective and radiative regions (i.e., "tropopause")
corresponds typically to a temperature of $2000\mathrm{K}$.

However, the boundary is likely to be above the tropopause. Even in radiative regions, eddy diffusion carries the evaporated materials of icy planetesimals upward like in the homosphere extending above the tropopause on the present Earth \citep[e.g.][]{Chamberlain+87}. Also, pollution occurs even in the upper envelope because the temperature there is enough to evaporate ice. While only a small fraction of the mass of planetesimals evaporates in the upper envelope \citep{Podolak+88}, the envelope is itself tenuous in such a region, so that the upper envelope can be also enriched in heavy elements.
Thus, we regard the boundary between the two layers as a free parameter in this study. To do so, we introduce "homopause temperature", $T_\mathrm{h}$; the values of $T_\mathrm{h}$ that we use in this study are $300,~500,~1000,~\mathrm{and}~2000\mathrm{K}$.

Thermodynamic quantities of the envelope gas are calculated under the assumption of chemical equilibrium. Although chemical equilibrium may not be achieved
in the region of low temperature because chemical reactions proceed slowly, 
this problem is beyond the scope of this paper. The following 13 constituents are considered, namely,
$\mathrm{H}$, $\mathrm{He}$, $\mathrm{C}$, $\mathrm{O}$, $\mathrm{H}_2$,
$\mathrm{O}_2$, $\mathrm{CO}$, $\mathrm{CO}_2$, $\mathrm{H}_2\mathrm{O}$,
$\mathrm{CH}_4$, $\mathrm{H}^+$, $\mathrm{O}^-$, and $\mathrm{e}^-$.
Thermodynamic quantities are calculated from the thermodynamic potentials.
The relevant physical quantities are given in NIST-JANAF thermochemical tables \citep{Chase98,Ott+00}.

We determine the mass fractions of carbon ($Z_\mathrm{h}^\mathrm{C}$)
and oxygen ($Z_\mathrm{h}^\mathrm{O}$) in the lower envelope in such a manner that
the sum of the two fractions is equal to $Z_\mathrm{h}$ and the ratio is conserved:
	\begin{equation}
		\begin{split}
		 Z_\mathrm{h} &= Z^\mathrm{C}_\mathrm{h} + Z^\mathrm{O}_\mathrm{h},  \\
		 Z^{\mathrm{C}}_\mathrm{h} &= 
		 	(1-\epsilon) Z^\mathrm{C}_\mathrm{A} 
			+ \epsilon Z^\mathrm{C}_\mathrm{B},  \\
		 Z^{\mathrm{O}}_\mathrm{h} &= 
		 	(1-\epsilon) Z^\mathrm{O}_\mathrm{A} 
			+ \epsilon Z^\mathrm{O}_\mathrm{B} ,
		\label{Zh}
		\end{split}
	\end{equation}
where $\epsilon$ is the mixing ratio of component B.
The mass fractions of hydrogen and helium in the mixture
are described by $\epsilon$ as well:
	\begin{equation}
		\begin{split}
		 X_\mathrm{h} &= 
		 	(1-\epsilon) X_\mathrm{A} + \epsilon X_\mathrm{B}, \\
		 Y_\mathrm{h} &= 
		 	(1-\epsilon) Y_\mathrm{A} + \epsilon Y_\mathrm{B}.
		\label{Xh_Yh}
		\end{split}
	\end{equation}
The ratio of $Z^\mathrm{C}_\mathrm{A}$ to $Z^\mathrm{O}_\mathrm{A}$ is
$0.004:0.011$, while that of $Z^\mathrm{C}_\mathrm{B}$ to $Z^\mathrm{O}_\mathrm{B}$ is $0.24:0.69$. In this study, we deal with $Z_\mathrm{h}$, which is equivalent to $\epsilon$, as a free parameter for simplicity because the actual value of $Z_\mathrm{h}$ depends on various processes such as ablation efficiency of icy planetesimals and their entry velocities and should be determined in a complicated manner.

The gas opacity is derived from opacity tables with different values of $X$ and
$Z$ given by J. Ferguson \citep{Alexander+94}. Their calculations include
other heavy elements besides $\mathrm{H},~\mathrm{He},~\mathrm{C},~
\mathrm{and}~\mathrm{O}$. All the elements
except $\mathrm{H}$ and $\mathrm{He}$ are assumed to be in
the solar abundances. This is reasonable because the abundances of
heavy elements in comet Halley are known to be similar to the solar abundances \citep{Mumma+93}.
The opacity tables of \citet{Alexander+94} are provided as a function of
$\log Q = \rho/T^3_6$, where $\rho$ is the density in $\mathrm{g~cm}^{-3}$ and $T_6$ the temperature in million K. 
If $\log Q$ exceeds the available range of the opacity tables, namely, $\log Q > 3$,
we calculate the gas opacity by extrapolating the tabular data.
Even if the values at $\log Q = 3$ are used for $\log Q > 3$ without extrapolation, no
significant difference in the critical core mass is found.

The Rosseland mean opacity of dust grains is calculated from
the monochromatic opacities of dust grains in protoplanetary discs presented by \citet{Semenov+03}.
The dust constituents, their evaporation
temperatures, and the size distribution of the dust grains used in
\citet{Semenov+03} are the same as those in \citet{Pollack+94}, while their optical constants refer to \citet{Henning+96}.
The difference between the grain opacities is nevertheless small \citep{Semenov+03}.
We introduce a factor, $f$, that represents reduction or enhancement
of the grain opacity, namely, grain opacities being $f$ times opacities of grains in disc gas.
The reasons why we introduce the factor are as follows.
When icy planetesimals go through the upper envelope, if undifferentiated
like comet Halley \citep[e.g.][]{Mumma+93}, they release dust grains embedded in a matrix of
ice upon evaporation. The deposited grains raise the grain
opacities in the upper envelope. On the other hand, opacities of dust grains in the envelope may be reduced due to
their coagulation and settling as described in Introduction \citep{Podolak03,Movshovitz+08,Movshovitz+10}.
In this study, the total opacity is expressed by
	\begin{equation}
		\kappa = f\kappa_\mathrm{gr} + \kappa_\mathrm{gas},
		\label{opc}
	\end{equation}
where $\kappa_\mathrm{gas}$ is the Rosseland mean opacity of gas and 
$\kappa_\mathrm{gr}$ is that of dust grains in protoplanetary discs.
\section{Critical Core Masses \label{sec.3}}
\vspace{0.2cm}

\begin{figure}
\includegraphics[width=0.8\hsize,clip]{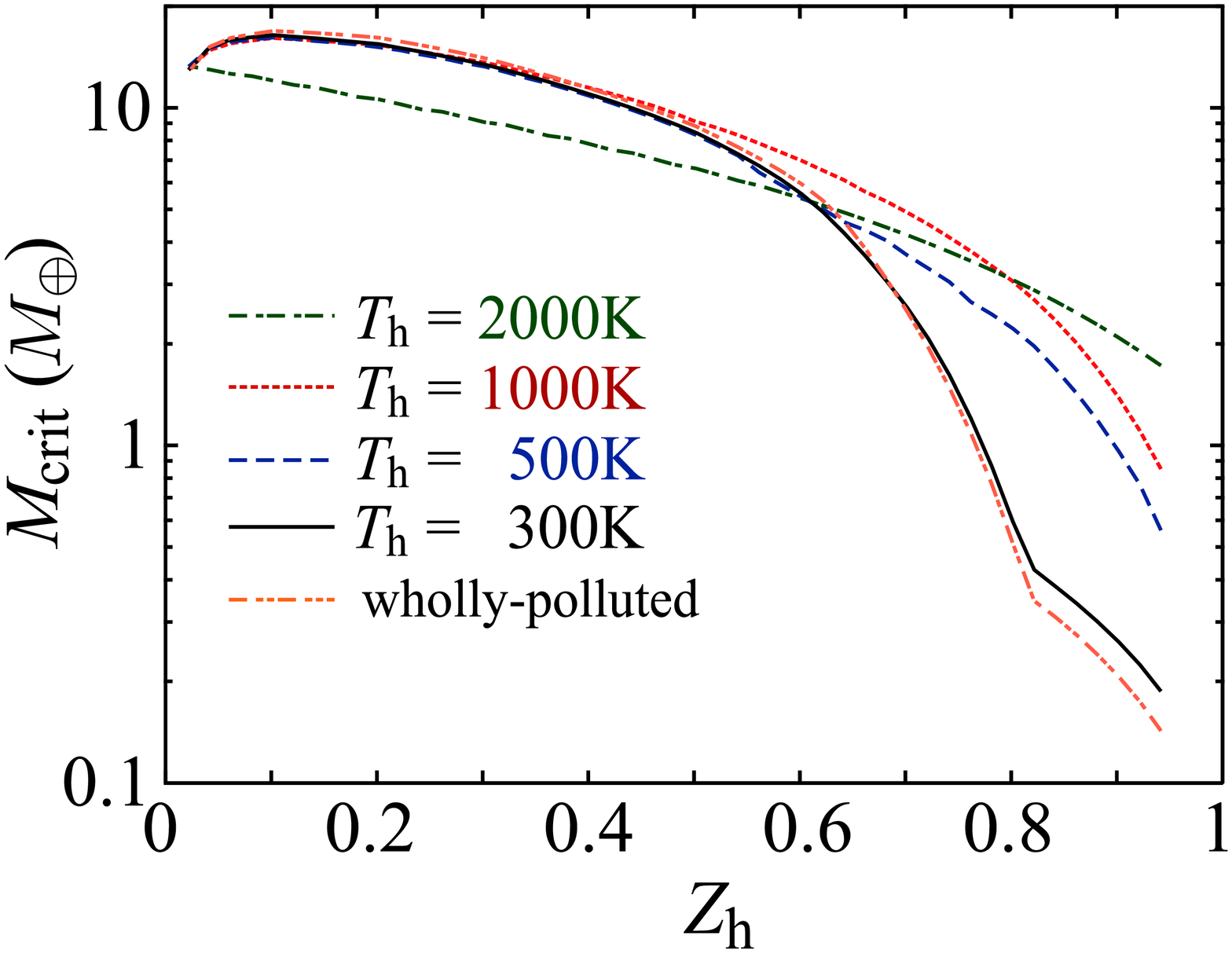}
\caption{The critical core mass ($M_\mathrm{crit}$)
as a function of the mass fraction of heavy
elements in the lower layer of the envelope, $Z_\mathrm{h}$.
The solid, dashed, dotted, and dot-dashed curves
correspond to the results for
$T_\mathrm{h} = 300,~500,~1000,~\mathrm{and}~2000\mathrm{K}$,
respectively.
The double dot-dashed curve is the result of wholly-polluted envelopes.
In all the calculations, we assume $L = 1 \times 10^{27}~\mathrm{erg/s}$ and $f = 1$.
A sudden change in the slope at $Z_\mathrm{h} = 0.815~\mathrm{for}~
T_\mathrm{h} = 300\mathrm{K}$ and the wholly-polluted envelopes
appears because
one of the major molecules, $\mathrm{H}_2$, is replaced with $\mathrm{CO}_2$ (see the text).
\label{fig2}}
\end{figure}

\begin{table}
\caption{Free Parameters and Their Values.\label{tbl1}}
\begin{flushleft}
\begin{tabular}{lrl}
\hline
Parameter & &  Value \\ 
\hline
homopause temperature, $T_\mathrm{h}$ & & 300, 500, 1000, 2000K \\
mass fraction of heavy elements & & \\
in the lower layer, $Z_\mathrm{h}$ & & 
$0.015-0.94$ \\
grain-depletion factor, $f$ & & $0, 0.01, 1, 10$ \\
luminosity, $L$ & & $1 \times 10^{26}, 10^{27}, 10^{28}, 10^{29} \mathrm{erg/s}$ \\
\hline
\end{tabular}
\end{flushleft}
\end{table}

Critical core masses are found in the same way as \citet{Mizuno80}.
For a set of four parameters listed in Table \ref{tbl1},
we determine the static structure of the envelope (i.e., $dS/dt = 0$) to find the core mass for a given protoplanetary total mass, $M_\mathrm{p}$. We increase $M_\mathrm{p}$ and
repeat the same procedure until the core mass reaches a first maximum, which is the critical core mass. In this section, we demonstrate how envelope pollution affects the critical core mass ($M_\mathrm{crit}$). Our results show that the heavy-element enrichment, in general, lowers $M_\mathrm{crit}$ and this behaviour of $M_\mathrm{crit}$ holds good for any choice of four parameters.

\subsection{Effects of Envelope Pollution by Icy Planetesimals \label{sec.3-1}}
\vspace{1em}

Figure \ref{fig2} shows $M_\mathrm{crit}$ as a function of the mass fraction
of heavy elements in the lower envelope, $Z_\mathrm{h}$, for four different homopause temperatures, $T_\mathrm{h}$.
The solid, dashed, dotted, and dot-dashed curves
correspond to $T_\mathrm{h} =300,~500,~1000,~\mathrm{and}$
\onecolumn
\begin{figure}
\centering
\includegraphics[width=0.7\hsize,clip]{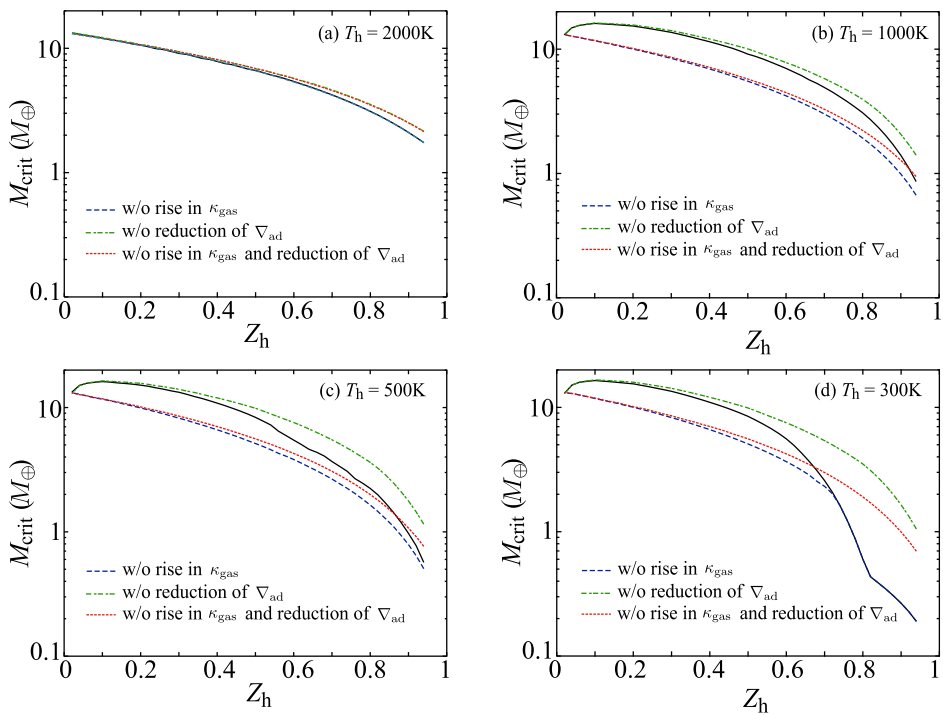}
\caption{The critical core mass ($M_\mathrm{crit}$)
as a function of $Z_\mathrm{h}$ for four different $T_\mathrm{h}$:
(a) $T_\mathrm{h} = 2000\mathrm{K}$,
(b) $T_\mathrm{h} = 1000\mathrm{K}$,
(c) $T_\mathrm{h} = 500\mathrm{K}$,
(d) $T_\mathrm{h} = 300\mathrm{K}$.
The solid curves are the same results as Fig.2.
The dashed and dot-dashed ones
exclude the rise in $\kappa_\mathrm{gas}$ and
the reduction of $\nabla_\mathrm{ad}$, respectively, and
the dotted one includes only the increase in $\mu$.
\label{fig3}}
\end{figure}
\begin{figure}
\centering
\includegraphics[width=0.7\hsize,clip]{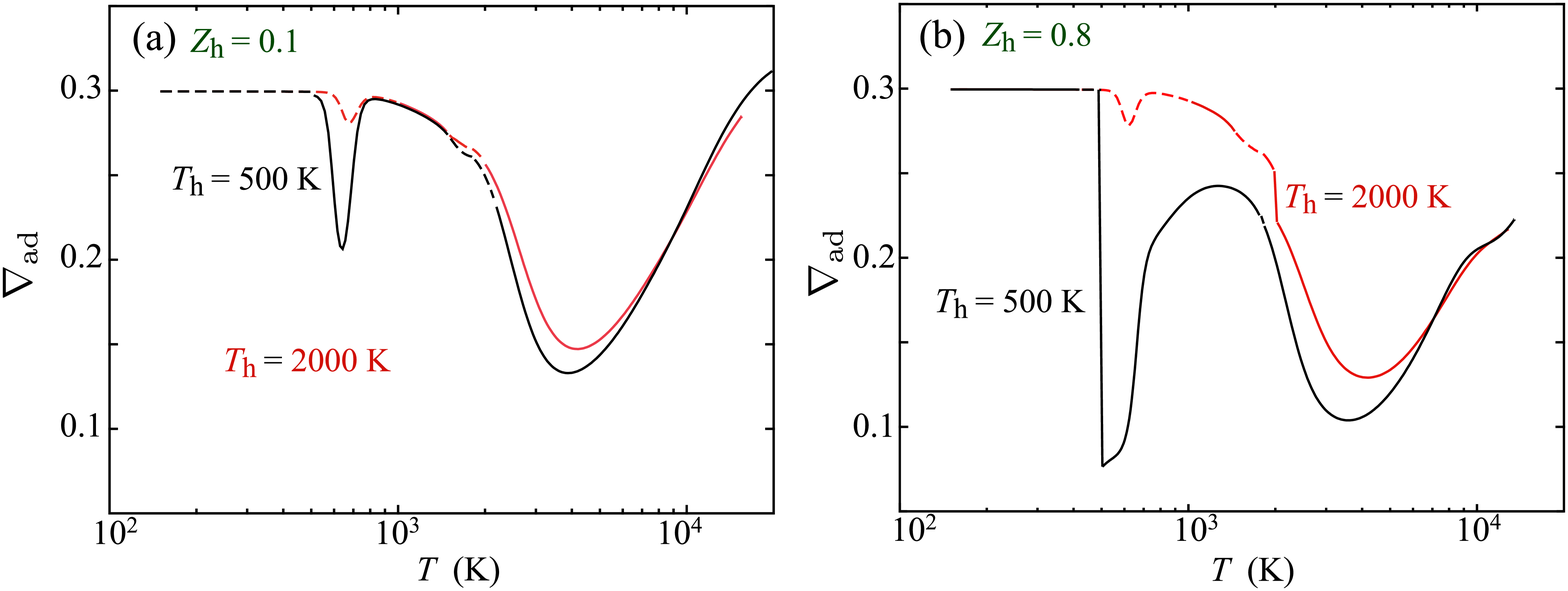}
\caption{The adiabatic temperature gradient ($\nabla_\mathrm{ad}$)
as a function of temperature
when $T_\mathrm{h} = 500\mathrm{K}$ and $T_\mathrm{h} = 2000\mathrm{K}$.
Convective and radiative regions are shown by solid and dashed lines, respectively.
The left panel plots the results for $Z_\mathrm{h} = 0.1$ and
the right panel plots those for $Z_\mathrm{h} = 0.8$
at $M_\mathrm{crit}$.
\label{fig4}}
\end{figure}
\begin{figure}
\centering
\includegraphics[width=0.7\hsize,clip]{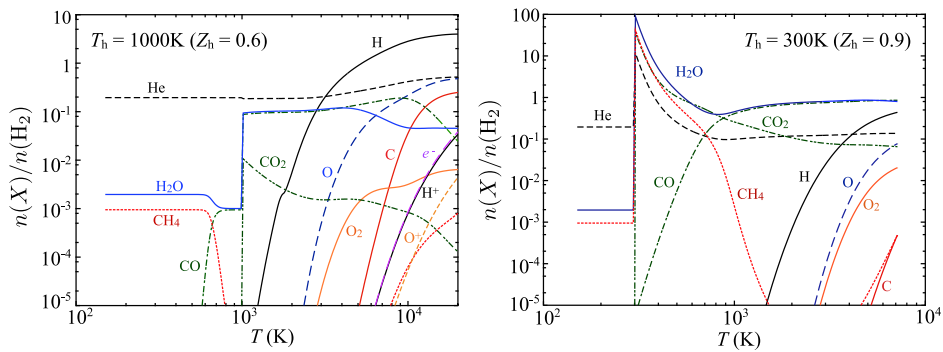}
\caption{The number density of chemical species, $n(\mathrm{X})$, relative
to that of molecular hydrogen, $n(\mathrm{H}_2)$, 
as a function of temperature
when $M_\mathrm{core} =$ $M_\mathrm{crit}$ for 
$T_\mathrm{h} = 1000~\mathrm{K}$ and $Z_\mathrm{h} = 0.6$ (the left panel) 
and $M_\mathrm{core} =$ $M_\mathrm{crit}$ for $T_\mathrm{h} = 300~\mathrm{K}$ 
and $Z_\mathrm{h} = 0.9$ (the right one).
\label{fig5}}
\end{figure}
\twocolumn$2000\mathrm{K}$, respectively.
We also examine the wholly polluted envelope
as an extreme case, which is shown by the double dot-dashed curve.
In all the calculations shown in Figure \ref{fig2}, we assume $L = 1 \times 10^{27}~\mathrm{erg/s}$ and $f = 1$.

As found in Fig.~\ref{fig2}, except for small increases for $Z_\mathrm{h} \leq 0.1$, 
the heavy-element enrichment lowers $M_\mathrm{crit}$ for any $T_\mathrm{h}$.
In particular, the reduction in $M_\mathrm{crit}$ is significant for high $Z_\mathrm{h}$.
In Fig.\ref{fig2}, one notices that $M_\mathrm{crit}$ depends on
$Z_\mathrm{h}$ and $T_\mathrm{h}$ in complicated manners.
There are three factors to change $M_\mathrm{crit}$ as described in Introduction.
To see how each effect contributes to change $M_\mathrm{crit}$,
we have done the following sensitivity tests; the results are presented in Fig.\ref{fig3}.
The solid curves correspond to the results shown in Fig.\ref{fig2}.
Other three curves represent the results for cases where one or two out of the three factors are artificially excluded.
\begin{itemize} 
\item The dashed curves: We have used $\kappa_\mathrm{gas}$ with the solar abundances throughout the envelope instead of including the rise in $\kappa_\mathrm{gas}$.
The difference between the solid and the dashed curves represents
the increment of $M_\mathrm{crit}$ caused by the rise in $\kappa_\mathrm{gas}$. \\
\item The dot-dashed curves: We have used $\nabla_\mathrm{ad}$ with the solar abundances throughout the envelope
to exclude the effect of reduction of $\nabla_\mathrm{ad}$.
The difference between the solid and the dot-dashed curves represents
the decrement of $M_\mathrm{crit}$ due to reduction of $\nabla_\mathrm{ad}$.\\
\item The dotted curves: We have used both $\kappa_\mathrm{gas}$ and
$\nabla_\mathrm{ad}$ with the solar abundances throughout the envelope
to extract only the effect of the increase in $\mu$.
The difference between the values of $M_\mathrm{crit}$ for $Z_\mathrm{h}=Z_\mathrm{A} (= 0.015)$ and the dotted curves shows the decrement of $M_\mathrm{crit}$ due to the increase in $\mu$.
\end{itemize}

First, in the case of $T_\mathrm{h} = 2000\mathrm{K}$,
the increase in $\mu$ is the most important
for lowering $M_\mathrm{crit}$, as shown in Fig.\ref{fig3}a.
The lower envelope ($T \geq 2000\mathrm{K}$)
is fully convective (see Fig.\ref{fig4}),
so that the rise in $\kappa_\mathrm{gas}$ has no influence on
the change in $M_\mathrm{crit}$.
That is why the heavy-element enrichment in the region of $T \geq 2000\mathrm{K}$
always decreases $M_\mathrm{crit}$ with increasing $Z_\mathrm{h}$.
As for all the other cases, $M_\mathrm{crit}$ increases at first and decreases
beyond $Z_\mathrm{h} = 0.1$.
When $Z_\mathrm{h} < 0.1$,
the increase in $\mu$ and reduction of $\nabla_\mathrm{ad}$ are inefficient in
lowering $M_\mathrm{crit}$ as shown in Fig.\ref{fig3}b-\ref{fig3}d, so that
the rise in $\kappa_\mathrm{gas}$ raises $M_\mathrm{crit}$.

For low $T_\mathrm{h}$ and relatively high $Z_\mathrm{h}$,
reduction of $\nabla_\mathrm{ad}$ is the most effective
in lowering $M_\mathrm{crit}$, 
as shown in Fig.\ref{fig3}d.
To understand this behaviour, 
we present profiles of $\nabla_\mathrm{ad}$ in the envelope
when $M_\mathrm{core} = M_\mathrm{crit}$ for
$T_\mathrm{h} = 500\mathrm{K}~\mathrm{and}~2000\mathrm{K}$ in Fig.\ref{fig4}:
$Z_\mathrm{h} = 0.1$ (left panel) and $Z_\mathrm{h} = 0.8$ (right panel).
The other $T_\mathrm{h}$ cases are not presented here because
the results for $T_\mathrm{h} = 300\mathrm{K}~\mathrm{and}~1000\mathrm{K}$
are quite similar to those for $T_\mathrm{h} = 500\mathrm{K}~\mathrm{and}
~2000\mathrm{K}$, respectively.
Convective and radiative regions are indicated by the solid and the dashed lines
in Fig.\ref{fig4}, respectively.
Two deep valleys of $\nabla_\mathrm{ad}$ are seen around $500-900\mathrm{K}$
and above $2000\mathrm{K}$ in Fig.\ref{fig3}.
Especially, $\nabla_\mathrm{ad}$ for 500-900K is much smaller in the case of $T_\mathrm{h} = 500\mathrm{K}$
than in the case of $T_\mathrm{h} = 2000\mathrm{K}$. 
Such small $\nabla_\mathrm{ad}$ is responsible for the small value of $M_\mathrm{crit}$.

The reason for the decrease in $\nabla_\mathrm{ad}$ is that several chemical reactions occur in that temperature range.
We present the number density of each chemical species, $n(\mathrm{X})$,
relative to that of molecular hydrogen, $n(\mathrm{H}_2)$, for two typical cases in Fig.\ref{fig5}:
$Z_\mathrm{h} = 0.6$ and $T_\mathrm{h} = 1000$K (left panel)
and $Z_\mathrm{h} = 0.9$ and $T_\mathrm{h} = 300$K (right panel).
Note that a sudden change in the slope at $Z_\mathrm{h} = 0.815$ of the solid line in Fig.~\ref{fig3}d appears because 
one of major molecules, $\mathrm{H}_2$,
is replaced with $\mathrm{CO}_2$ at $Z_\mathrm{h} \geq 0.815$
as shown in Fig.\ref{fig5}b.
In the regime of $400-900\mathrm{K}$,
there occur the abrupt decreases in $\mathrm{H}_2\mathrm{O}$ and $\mathrm{CH}_4$, which are stable chemical species at low temperatures, and
the sharp increases in $\mathrm{CO}$ and $\mathrm{CO}_2$.
Those chemical reactions have great impacts on reduction of $\nabla_\mathrm{ad}$.
Above $2000\mathrm{K}$, dissociations of several kinds of molecules such as $\mathrm{H}_2$, $\mathrm{H}_2\mathrm{O}$, and $\mathrm{CO}_2$ take place and
reduce $\nabla_\mathrm{ad}$.
As a result, the envelope is rich in $\mathrm{CO}$, $\mathrm{H}$,
and $\mathrm{O}$ above $2000\mathrm{K}$.

\subsection{Dependence on grain opacity \label{sec.3-2}}
\vspace{1em}

\begin{figure}
\includegraphics[width=0.8\hsize,clip]{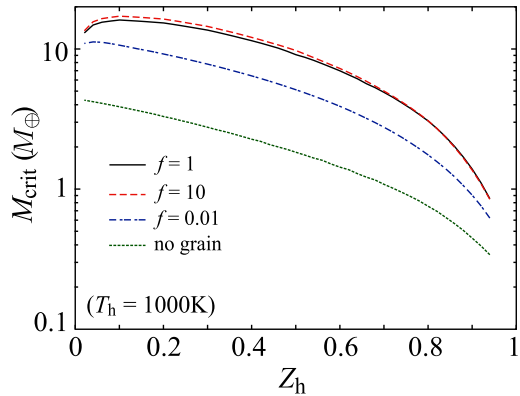}
\caption{Critical core masses ($M_\mathrm{crit}$)
for four different $f$ in the case
of $T_\mathrm{h} = 1000~\mathrm{K}$.
The solid, dashed, dot-dashed, and dotted curves
correspond to $f = 1,~10,~0.01,~0$ (no grain).
In all of the calculations, $L = 1 \times 10^{27}~\mathrm{erg/s}$ is assumed.
Only for the results of $f = 0$ are given by \citet{Hori+10}.
The solid curve is the same as the result represented by the dotted one in Fig.2.
\label{fig6}}
\end{figure}

We considered only the case of $f = 1$ in Section 3.1.
The upper layer of the envelope may, however, be grain-rich or grain-poor,
as described in Section 2.
We consider both grain-rich ($f > 1$) and grain-poor ($f < 1$) envelopes in this subsection.
The upper layer (i.e., $T < T_\mathrm{h}$) is convectively stable in most cases.
Thus, it is to be verified that the spatial distribution of added dust grains
in that layer is homogeneous;
nevertheless, we adopt a constant value of $f$ throughout the upper layer for simplicity.

The behaviours of $M_\mathrm{crit}$ for $T_\mathrm{h} = 1000\mathrm{K}$ with respect to
four different $f$ are shown in Fig.\ref{fig6}.
The solid curve is the same as that in Fig.\ref{fig2}b.
The dashed, dot-dashed, and dotted curves
correspond to $f = 10,~0.01,~\mathrm{and}~0$, respectively.
Results for $f > 10$ hardly differ from that for $f =10$.
In all the calculations, $L = 1 \times 10^{27}~\mathrm{erg/s}$ is assumed. 

As shown in Fig.~\ref{fig6}, $M_\mathrm{crit}$ increases as a whole (i.e., for any $Z_\mathrm{h}$), as $f$ increases. This is because dust grains are the dominant opacity source in the upper layer. The increase in $M_\mathrm{crit}$ is found to level off around $f =1$. This is because the part of the envelope where the grain opacity dominates becomes convective; on the other hand, the slight increase for $Z_\mathrm{h} \lesssim 0.1$ is due to increase in gas opacity in a relatively high-temperature, radiative region in the layer where the gas opacity dominates.
We have also investigated the behaviour of $M_\mathrm{crit}$ for different $T_\mathrm{h}$
and confirmed that both cases of grain-rich and grain-poor envelopes show trends
similar to that for $T_\mathrm{h} = 1000\mathrm{K}$.

\subsection{Dependence on luminosity \label{sec.3-3}}
\vspace{1em}

\begin{table}
\caption{Critical core masses for $T_\mathrm{h} = 1000~\mathrm{K}$
and $f = 1$, but four different $L~(\mathrm{erg/s})$ .\label{tbl3}}
\begin{flushleft}
\begin{tabular}{lrrrr}
\hline
 $Z_\mathrm{h}$ & $1\times10^{26}$
 & $1\times10^{27}$ & $1\times10^{28}$ & $1\times10^{29}$ \\
\hline
0.1 & $9.8M_\oplus$  & $16M_\oplus$ & $27M_\oplus$ & $43M_\oplus$ \\
0.3 & $8.3M_\oplus$  & $14M_\oplus$ & $22M_\oplus$ & $35M_\oplus$ \\
0.6 & $4.5M_\oplus$  & $7.0M_\oplus$ & $11M_\oplus$ & $16M_\oplus$ \\
0.9 & $1.0M_\oplus$ & $1.4M_\oplus$ & $1.8M_\oplus$ & $1.8M_\oplus$ \\
\hline
\end{tabular}
\end{flushleft}
\end{table}

We assumed $L = 1 \times 10^{27}~\mathrm{erg/s}$ in all the calculations
shown above. The dependence of $M_\mathrm{crit}$ on $L$ is qualitatively the same as that in the case of solar-composition envelopes \citep[][]{Ikoma+00}. Here we quantify the dependence for polluted envelopes. We calculate $M_\mathrm{crit}$ for three different values of $L = 1\times 10^{26}$, $1\times 10^{28}$, and $1\times 10^{29}~\mathrm{erg/s}$.
Those values are chosen based on accretion rates of planetesimals that formation theory predicts \citep{Pollack+96,Fortier+09}.
The results are listed in Table \ref{tbl3}, where $T_\mathrm{h} = 1000\mathrm{K}$.
Table \ref{tbl3} shows that the decreasing trend of $M_\mathrm{crit}$ with $Z_\mathrm{h}$ is the same, irrespective of $L$.
We also have found that this is true for other $T_\mathrm{h}$.

\subsection{Dependence on semimajor axis \label{sec.3-4}}
\vspace{1em}

We discussed $M_\mathrm{crit}$ at the present location of Jupiter
in the previous subsections.
In this subsection, we consider $M_\mathrm{crit}$ at two different semimajor axes,
1AU and 10AU. 
The choice of the two values is based on recent theories of the snow line in protoplanetary discs. Many relevant studies suggest that the snow line moves and reaches $\sim 1\mathrm{AU}$ \citep{Sasselov+00,Davis05,Garaud+07,Min+11}.
We thus adopt 1AU as the innermost orbit of the protoplanet that
can experience envelope pollution by icy planetesimals.
The disc density ($\rho_\mathrm{disc}$) and temperature ($T_\mathrm{disc}$),
namely, the outer boundary conditions,
are taken as mentioned in Section 2.

The sensitivity of $M_\mathrm{crit}$ to
the outer boundary conditions was already investigated in detail
\citep{Mizuno80,Stevenson82,Wuchterl93,Ikoma+01,Rafikov06}:
It is known that
$M_\mathrm{crit}$ for wholly radiative envelopes is almost independent of
the outer boundary conditions, while
$M_\mathrm{crit}$ for fully convective envelopes depends significantly on them.
Polluted envelopes tend to be convective, especially near the central star. 
The critical disc density above which the envelope is wholly convective is given by equation~(13) of \citet{Ikoma+01}: 
	\begin{align}
			\frac{\rho^\mathrm{crit}_\mathrm{disc}}{\rho_\mathrm{MMSN}}
			&\sim 5 \left(\frac{a_\mathrm{p}}{1\mathrm{AU}}\right)^{3/2}
			\left(\frac{T_\mathrm{disc}}{T_\mathrm{MMSN}}\right)^{5/2}
			\left(\frac{M_\mathrm{c}}{10M_\oplus} \right)
			\left(\frac{T_\mathrm{evap}}{1500\mathrm{K}}\right)^{1/2}
			\nonumber \\
			&\times \left(\frac{L}{1\times10^{27}~\mathrm{erg/s}}\right)^{-1}
			\left(\frac{\kappa_\mathrm{evap}}{1~\mathrm{cm}^2\mathrm{/g}}\right)^{-1}
			\left(\frac{\mu}{2.3} \right),
		\label{rho_crit}
	\end{align}
where $T_\mathrm{evap}$ is the evaporation temperature of grains,
$\kappa_\mathrm{evap}$ is the grain opacity near $T = T_\mathrm{evap}$, and
the subscript "MMSN" means that the quantity takes the value of the MMSN model \citep{Hayashi81}.
Based on the fact that
$\kappa_\mathrm{evap} \sim 30-100~\mathrm{cm}^2~\mathrm{g}^{-1}$
for $T_\mathrm{h} = 300\mathrm{K}$ and
$\kappa_\mathrm{evap} \sim 3-5~\mathrm{cm}^2~\mathrm{g}^{-1}$
for other $T_\mathrm{h}$ at $T_\mathrm{evap} \sim 1000\mathrm{K}$,
$\rho_\mathrm{disc} > \rho^\mathrm{crit}_\mathrm{disc}$ is satisfied for all the models of
$1\mathrm{AU}$ and $T_\mathrm{h} = 300\mathrm{K}$.
In fact, the outer envelope becomes fully convective
in the cases of $T_\mathrm{h} = 300\mathrm{K}$ and $a_\mathrm{p} = 1\mathrm{AU}$. 
It is, thus, worth knowing the impact of outer boundary conditions on $M_\mathrm{crit}$ in the case of polluted envelopes.

\begin{figure}
\includegraphics[width=0.8\hsize,clip]{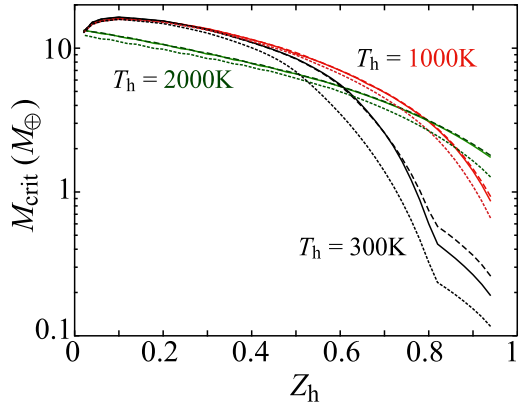}
\caption{Critical core masses ($M_\mathrm{crit}$) as a function of $Z_\mathrm{h}$
for $T_\mathrm{h} = 300\mathrm{K}, 1000\mathrm{K},~\mathrm{and}~2000\mathrm{K}$.
The same as Fig.2, but for different semimajor axes.
The solid, dashed, and dotted curves adopt the outer boundary conditions of
$5.2\mathrm{AU}$, $10\mathrm{AU}$, and $1\mathrm{AU}$, respectively. 
\label{fig7}}
\end{figure}

Figure \ref{fig7} shows $M_\mathrm{crit}$ as a function of $Z_\mathrm{h}$ at $5.2\mathrm{AU}$ (solid line), $1\mathrm{AU}$~(dotted one) and $10\mathrm{AU}$~(dashed one) for $T_\mathrm{h} = 300,~1000$, and $2000\mathrm{K}$; the values of the other parameters are the same as in Fig.\ref{fig2}. The trend of $M_\mathrm{crit}$ is the same for all the cases. The impact of outer boundary conditions is small at $a_\mathrm{p} > 1~\mathrm{AU}$; $M_\mathrm{crit}$ differs at most by a factor of 2 even for $T_\mathrm{h} = 300~\mathrm{K}$ and $Z_\mathrm{h} \sim 1$. 

\section{Gas Accretion Time-scale \label{sec.4}}

As mentioned in Introduction, gas accretion is slower for smaller core mass, if the other parameters being the same. We need to check the time-scale for gas accretion in addition to the critical core mass.
In this section, we demonstrate that envelope pollution by planetesimals
can also hasten the gas accretion on to a protoplanet.
We adopt the growth time of the envelope, $\tau_\mathrm{gas}$, defined in \citet{Ikoma+00} as the typical time-scale of gas accretion on to the protoplanet
(see also \citet[][]{Ikoma+06b,Hori+10}).
We consider the protoplanet with a given core mass.
We simulate accumulation of disc gas after the core growth stops.
Both $Z_\mathrm{h}$ and $T_\mathrm{h}$  in those simulations
are assumed to be constant in time.
We discuss the validity of this assumption in Section 5.

\begin{figure}
\includegraphics[width=0.8\hsize,clip]{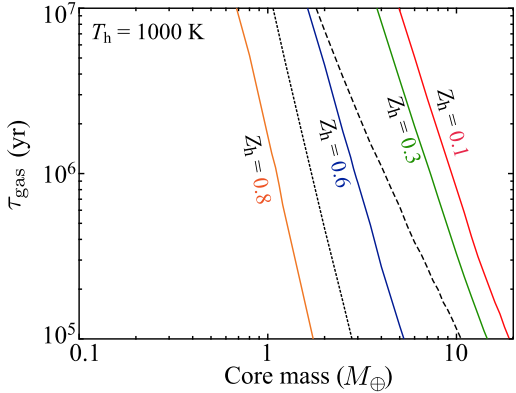}
\caption{The growth time-scale of the envelope ($\tau_\mathrm{gas}$) as a function of core masses. Four solid lines represent results of four different $Z_\mathrm{h}$
for $T_\mathrm{h} = 1000\mathrm{K}$ and $f = 1$:
$Z_\mathrm{h} = 0.1$, $Z_\mathrm{h} = 0.3$,
$Z_\mathrm{h} = 0.6$, and $Z_\mathrm{h} = 0.8$.
For comparison, we plot the results of $f = 0.01$ with solar abundances (dashed line)
and grain-free envelopes with the solar abundances, $f = 0$ (dotted one),
which are found by \citet{Hori+10}. 
\label{fig8}}
\end{figure}

Figure \ref{fig8} shows $\tau_\mathrm{gas}$ as a function of the core mass for
$T_\mathrm{h} = 1000\mathrm{K}$, $f = 1$, and $Z_\mathrm{h} = 0.1,~0.3,~0.6,~\mathrm{and}~0.8$ (solid lines).
For comparison, we plot the results for envelopes with the solar abundances: $f = 0.01$  (dashed line) and grain-free ($f = 0$) envelopes (dotted one).
This figure demonstrates that envelope pollution by icy planetesimals
accelerates gas accretion on to the protoplanet significantly.  
The acceleration is more effective
compared to that by reduction of grain opacities in the envelope.
We do not suppose reduction of grain opacities in the upper envelope in those simulations.
In addition to envelope pollution, if we take into account reduction of grain opacities,
it gives positive feedback to shorten $\tau_\mathrm{gas}$
(see Table \ref{tbl4}).

\begin{figure}
\includegraphics[width=0.8\hsize,clip]{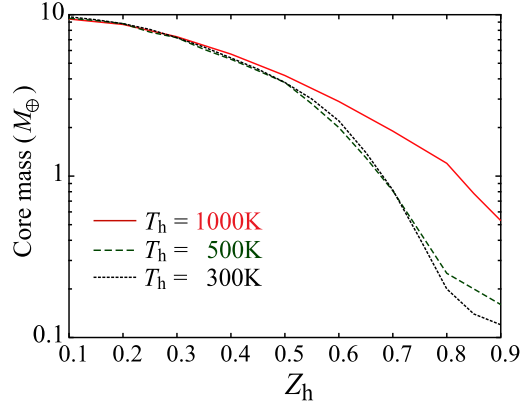}
\caption{Core masses with $\tau_\mathrm{gas} = 1$Myr for three $T_\mathrm{h}$
: $T_\mathrm{h} = 1000\mathrm{K}$ (solid),
$500\mathrm{K}$ (dashed),
$300\mathrm{K}$ (dotted).
In all the calculations, we assume $f = 1$.
\label{fig9}}
\end{figure}

Gas giant formation must be finished before disc gas disappears.
The disc lifetime is believed to be 1-10 Myrs based on the disc frequency estimated from near-IR excess ($JHKL$-band) observations of stars in young clusters \citep{Haisch+01,Hernandez+07}.
We now consider the core masses with $\tau_\mathrm{gas} \leq1\mathrm{Myr}$
as allowable for gas giant formation.
Minimum core masses ($M^\mathrm{min}_\mathrm{core}$)
that satisfy the criterion ($\tau_\mathrm{gas} = 1\mathrm{Myr}$)
are listed in Table \ref{tbl4} for $f = 1$ and $f = 0.01$.
We cannot simulate quasi-static evolution of the protoplanet's envelope with smaller cores than $0.1M_\oplus$ because of numerical difficulty.
In such a case, we show $M^\mathrm{min}_\mathrm{core} < 0.1M_\oplus$ in Table \ref{tbl4}.
In Fig.\ref{fig9}, we also plot $M^\mathrm{min}_\mathrm{core}$ as a function of $Z_\mathrm{h}$ 
for three $T_\mathrm{h}$, $1000\mathrm{K}$ (solid),
$500\mathrm{K}$ (dashed), and $300\mathrm{K}$ (dotted).
The behaviours of $M^\mathrm{min}_\mathrm{core}$ with respect of $Z_\mathrm{h}$ are quite similar to those of $M_\mathrm{crit}$.
We find that envelope pollution has 
contributions to hasten $\tau_\mathrm{gas}$ in the similar manner to
the lowering of $M_\mathrm{crit}$.
Our results suggest that envelope pollution by icy planetesimals has the potential to make gas giant formation with small cores possible; for example, 
gas giants with cores smaller than $1M_\oplus$ can capture disc gas by 1Myr
when $Z_\mathrm{h} \geq 0.7$ for $f = 1$ or
when $Z_\mathrm{h} \geq 0.5$ for $f = 0.01$.

\section{Discussions \label{sec.5}}

\begin{table}
\caption{Core masses with $\tau_\mathrm{gas} = 1\mathrm{Myr}$.\label{tbl4}}
\begin{flushleft}
\begin{tabular}{lrrrrr}
\hline
$T_\mathrm{h} (\mathrm{K})$ & $f$ & $Z_\mathrm{h}$  = 0.1 & $Z_\mathrm{h}$  = 0.3
& $Z_\mathrm{h}$  = 0.5 & $Z_\mathrm{h}$  = 0.9 \\
\hline
300 & $1$ & $9.7M_\oplus$  & $7.2 M_\oplus$ & $3.8M_\oplus$ & $0.12M_\oplus$ \\
& $0.01$  & $4.8M_\oplus$  & $2.4M_\oplus$ & $ 0.92M_\oplus$  & $< 0.1M_\oplus$ \\
500 & $1$ & $9.5M_\oplus$  & $7.2 M_\oplus$ & $3.8M_\oplus$ & $0.16M_\oplus$ \\
& $0.01$  & $3.8M_\oplus$  & $2.0M_\oplus$ & $ 0.85M_\oplus$  & $< 0.1M_\oplus$ \\
1000 & $1$ & $9.4M_\oplus$  & $7.3 M_\oplus$ & $4.2M_\oplus$ & $0.53M_\oplus$ \\
& $0.01$  & $3.1M_\oplus$  & $1.9M_\oplus$ & $ 0.95M_\oplus$  & $< 0.1M_\oplus$ \\
\hline
\end{tabular}
\end{flushleft}
\end{table}

\subsection{Dilution during runaway gas accretion}
We assumed that both $Z_\mathrm{h}$ and $T_\mathrm{h}$ are constant with time. This may be oversimplification and questionable, especially in the phase of runaway gas accretion. When the critical core mass is attained, the accretion rate of disc gas is much higher than that of planetesimals. In an extreme case where the unpolluted outer envelope never exchanges material with the polluted lower envelope, the gas accretion results in increasing the mass only of the upper envelope. The lower envelope behaves like a part of the ``core''. In this case, the envelope pollution does not resolve the problem of the slow formation of gas giants with small cores.

In the other extreme case where the inner and outer envelopes exchange material instantaneously between each other via eddy diffusion, accreting fresh disc gas dilutes the polluted lower envelope, which results in decelerating the disc-gas accretion. On the other hand, the dilution is inevitably accompanied by mass growth of the envelope, which accelerates the disc-gas accretion. 

Unfortunately, it is uncertain whether or not mixing occurs effectively, as follows.
A characteristic time-scale of eddy diffusion, $\tau_\mathrm{eddy}$, is given by 
	\begin{equation}
		\tau_\mathrm{eddy} \sim \frac{H^2_p}{K_\mathit{zz}},
		\label{t_eddy}
	\end{equation}
where $H_p$ is the pressure scale-height and $K_\mathit{zz}$ is the coefficient of eddy diffusion. 
Estimated values of $\tau_\mathrm{eddy}$ are listed in Table 4, where 
we have used our numerical values of $H_p$ at the tropopause when $M_\mathrm{core} = M_\mathrm{crit}$. We have adopted $K_\mathit{zz} = 10^6-10^8~\mathrm{cm^2~s^{-1}}$ which correspond to values for the present Jupiter and Saturn \citep[e.g.][]{Chamberlain+87}, although $K_{zz}$ for protoplanetary envelopes is uncertain.
The lower and upper values of $\tau_\mathrm{eddy}$ at each $Z_\mathrm{h}$ and
$K_\mathit{zz}$ in Table 4 are $\tau_\mathrm{eddy}$ of $T_\mathrm{h} = 2000$K and
that of $T_\mathrm{h} = 300$K, respectively.
As seen in this table,
$\tau_\mathrm{eddy}$ ranges widely from $10^3$ to $10^8~\mathrm{yr}$. 
Since time-scales of gas accretion of interest are $10^6$--$10^7$ years (which correspond to the observed disc lifetime), we can derive no definite conclusion as to whether dilution occurs or not, because of the uncertainty of $\tau_\mathrm{eddy}$.  Mixing in accreting protoplanetary envelopes is found to be an important issue with gas giant formation.

\begin{table}
\caption{$\tau_\mathrm{eddy}$ at the critical core masses.\label{tbl5}}
\begin{flushleft}
\begin{tabular}{lrr}
\hline
$Z_\mathrm{h}$ & $K_\mathrm{zz} = 10^6~(\mathrm{cm}^2~\mathrm{s}^{-1})$
& $K_\mathrm{zz}  = 10^8~(\mathrm{cm}^2~\mathrm{s}^{-1})$ \\
\hline
0.1 & $6\times10^{6}$-$1\times10^{8}~\mathrm{yr}$
& $6\times10^{4}$-$1\times10^{6}~\mathrm{yr}$ \\
0.3 & $3\times10^{6}$-$6\times10^{7}~\mathrm{yr}$
& $3\times10^{4}$-$6\times10^{5}~\mathrm{yr}$ \\
0.6 & $3\times10^{5}$-$6\times10^{6}~\mathrm{yr}$
& $3\times10^{3}$-$6\times10^{4}~\mathrm{yr}$ \\
0.8 & $3\times10^{5}$-$1\times10^{6}~\mathrm{yr}$
&  $3\times10^3$-$1\times10^4~\mathrm{yr}$\\%
\hline
\end{tabular}
\end{flushleft}
\end{table}

\subsection{Non-ideality of envelope gas}
In this study, we considered the envelope gas as ideal. In the case of the envelope with the solar abundances, we find that non-ideal effects cause less than $10\%$ change in the critical core mass. This paper aims to point out the importance to incorporate the effects of heavy-element enrichments in simulations of the structure and evolution of protoplanetary envelopes and to sort out what has a great impact on the critical core mass. Including the effects of non-ideality of gas is a future study. We found that reactions between molecular compounds play a important role in determining the critical core mass. Thus, we need thermodynamic quantities of non-ideal mixture of at least H, He, C, and O.

\subsection{Dynamical stability}
We should also pay attention to the dynamical stability of the envelope.
\citet{Tajima+97} found that the envelope with solar abundances is stable
dynamically throughout quasi-static evolution up to Jupiter's mass.
However, it is not obvious whether the envelope polluted by heavy elements is stable or not.
In this study, reduction of $\nabla_\mathrm{ad}$ due to molecular dissociations
is one of important effects to lower $M_\mathrm{crit}$ and shorten $\tau_\mathrm{gas}$.
Vibrational instability due to molecular dissociation may excite nonlinear hydrodynamical waves by $\kappa$-mechanism because of reduction of $\nabla_\mathrm{ad}$ and
the rise in $\kappa_\mathrm{gas}$ \citep{Wuchterl91}. Nonlinear hydrodynamical waves may cause dynamical expansion of the envelope and the ejection of a large part of the envelope.
Thus, the dynamical stability of the polluted envelope should be also investigated in the
future.

\subsection{Implication for Jupiter and Saturn}
It is worth while to consider the origin of Jupiter and Saturn from the viewpoint of core masses.
Interior modellings of the two planets suggest that Jupiter's core is smaller than $10 M_\oplus$, while Saturn's core is larger than $10 M_\oplus$ \citep{Saumon+04,Nettelmann+08}. 
\citet{Militzer+08} reported that Jupiter may possess a massive core of $> 10 M_\oplus$, but their interior modellings with rigid rotation are inconsistent with the observed value of the 4th-order gravitational moment, $J_4$. 
In addition, \citet{Fortney+10} pointed out that the mass fraction of helium used by \citet{Militzer+08} is responsible for a massive core of Jupiter. 
Although the exact mass of Jupiter's core is still an open question,
the core mass of Jupiter suggested by interior modellings is, on average, smaller than that of Saturn.

In this study, we demonstrated that envelope pollution by icy planetesimals enables formation of gas giants with small cores, which seems compatible with recent modellings of Jupiter. If so, how about the origin of Saturn that is inferred to have a large core?
The large core inside Saturn may reflect the history of high accretion rates of planetesimals. 
A variety of ideas for high accretion rates have been investigated so far --- e.g., the influence of growing proto-Jupiter and the other protoplanets
\citep{Guilera+10}, Type I migration of protoplanets \citep{Alibert+05}, high stellar metallicity and a large initial disc mass \citep{Thommes+06},
high initial surface density of solid materials \citep{Dodson-Robinson+08,Rafikov11}.
However, high accretion rates of planetesimals could have a risk of forming Saturn with a small core. This is because a large amount of small-sized fragments are produced in the Saturn's region due to frequent high-velocity collisions of planetesimals, so that the envelope should be polluted heavily by icy planetesimals.
Thus, we require to revisit accretion rates of planetesimals on to proto-Saturn, for example, in the case where Jupiter first emerges and other planetary cores still continue to grow. This is because planetary accretion proceeds more rapidly in the inner regions, so that proto-Jupiter may have formed a small core first and triggered runaway gas accretion due to envelope pollution by icy planetesimals.

As one alternative idea for a large core of Saturn, \citet{Li+10} suggested that Saturn might have experienced an impact of a (proto-) gas giant and then merged. It is still an open question whether such an impact on Saturn happened in reality in a multiple-protoplanet system. Direct N-body simulations of competitive core growth beyond the snow line will give an answer to this problem. 

We should also comment on another possibility that the disc instability scenario may be able to explain gas giants with small cores. 
The disc instability scenario is usually a single-stage model in which a massive protoplanetary disc, typically $0.1 M_\odot$, becomes gravitationally unstable and rapidly collapses to form a proto-gas giant planet \citep[e.g.][]{Cameron78,Boss06,Mayer+07,Durisen+07,Boley09,Inutsuka+10}.
Recent works on the disc instability scenario showed that a gaseous clump is able to capture planetesimals in the course of its rapid contraction, assuming concurrent formation of large-sized planetesimals \citep{Helled+08a,Helled+08b,Helled+09} or a gaseous clump is formed in a dense region where solid materials (1$\mu\mathrm{m}$-10$\mathrm{cm}$) are collected aerodynamically by spiral arms \citep{Boley+11}.
In those scenarios, grains that grow to typically $\sim10~\mathrm{cm}$ can settle down toward the centre against the stirring of convective motion and form a relatively small core. 
Also the disc instability scenario is faced with the problem of Saturn formation,
although Jupiter with a small core could be formed by the disc instability.
One possibility to form a large core of Saturn is that the existence of Jupiter influences the distribution of planetesimals near Jupiter's region and enhances the core growth of Saturn. 
As mentioned in the above paragraph, this situation is the almost same as the envisioned picture of the Saturn formation in the core-accretion model taking account into envelope pollution. 
More recently, \citet{Nayakshin10} has proposed another idea of gas giant formation with large cores by the disc instability: Grain sedimentations inside a large gaseous clump yield a large core and trigger rapid gas accretion on to the core. 
When the clump migrates into a few AU from a parent star, it experiences tidal disruption and evaporation due to stellar irradiation. It is peeled off the outer metal-poor envelope of the clump and then results in a Saturn-like planet.
In the future, further detailed studies will verify his idea.

\subsection{The amount of heavy elements in envelopes of extrasolar gas giants}

Recent measurements of the mass and radius of transiting extrasolar gas giants (EGG) allow us to constrain planetary composition \citep[e.g.][]{Guillot08}. Those internal structures showed a variety of the total amount of heavy elements contained in transiting EGGs, including cores. Although it is not clear whether such a diversity of the amount of heavy elements in EGGs reflects a diversity of their core masses, core masses of EGGs are a key test to verify how efficiently our idea of envelope pollution works in reality. Further transit observations of EGGs and our comprehensive understanding of interior structures of EGGs will help us to reveal formation of giant planets.

\subsection{Implication for Uranus and Neptune}
Finally, we consider an implication of interior structures of Uranus and Neptune. 
Envelope pollution by icy planetesimals (and erosion of their icy shells) may be in good agreement with interior modellings of the present Uranus and Neptune.
The $\mathrm{H}/\mathrm{He}$ envelopes enriched by volatile molecules such as $\mathrm{H}_2\mathrm{O}$ and $\mathrm{CH}_4$ have been proposed in order to reproduce their observed gravitational moments ($J_2$ and $J_4$), which are poorly constrained \citep{Podolak+95,Marley+95}. 
Many recent works on interior modellings of Uranus and Neptune also suggested that they may have the $\mathrm{H}/\mathrm{He}$ envelopes into which "ices" are mixed instead of traditional three-layer models, that is density-stratified interiors \citep{Podolak+00,Helled+10}.
On the other hand, the fact that the present Uranus and Neptune have non-axisymmetric, non-dipolar magnetic fields may favour the existence of stably-stratified fluid layers beneath the outer thin shell that drives dynamo action \citep{Stanley+04,Stanley+06}. 
In addition, the compositional stratification in the interior may be responsible for extremely low-luminous Uranus \citep{Podolak+91,Hubbard+95}.
It is still unclear whether the H/He envelopes are gradually-mixed with ices toward the centres or not.
In any case, both Uranus and Neptune are likely to possess the outer polluted envelopes from the viewpoint of the gravitational harmonics and thermal evolution. 
Their polluted envelopes may be consistent with the picture of envelope pollution by incoming icy planetesimals, although we cannot rule out the another possibility for Uranus that a giant impact on a primitive Uranus results in the polluted $\mathrm{H}/\mathrm{He}$ envelope because the current Uranus has the extreme axial tilt \citep{Korycansky+90,Slattery+92}.

\vspace{-1em}
\section{Summary and Conclusions \label{sec.6}}
\vspace{1em}

We have investigated how envelope pollution by incoming icy planetesimals
affects critical core masses ($M_\mathrm{crit}$)
and growth time-scales of the envelope ($\tau_\mathrm{gas}$).
We considered that the envelope has two-layer structure:
the upper layer with the solar abundances (non-polluted layer)
and the lower one with mixture of solar and comet-Halley-like compositions
(polluted layer). We introduced two key parameters, 
the mass fraction of the heavy elements in the lower envelope
($Z_\mathrm{h}$) and the homopause temperature ($T_\mathrm{h}$) which defines the boundary between the two layers. The main results that we obtained in this study are summarized as follows.
\vspace{1em}

\begin{itemize}
\item[(1)~] Critical core masses:\\ \\
Envelope pollution by icy planetesimals lowers $M_\mathrm{crit}$ for any choice of values of $T_\mathrm{h}$ and most of the range of $Z_\mathrm{h} (\gtrsim 0.1)$  (Fig.\ref{fig2}).
Widely-polluted (low $T_\mathrm{h}$) and
highly-polluted (high $Z_\mathrm{h}$) envelopes lessen $M_\mathrm{crit}$ remarkably.
The increases in the molecular weight ($\mu$) and reduction of adiabatic temperature gradient ($\nabla_\mathrm{ad}$) are responsible for
the lowering of $M_\mathrm{crit}$ (Fig. \ref{fig3}).
In particular, remarkable reduction of $\nabla_\mathrm{ad}$ due to chemical reactions for 400-900K and molecule dissociations for $>$ 2000K (Figs. \ref{fig4} and \ref{fig5}).
Such behaviour of $M_\mathrm{crit}$ is the same for any choices of other parameters, the grain opacities (Fig.~\ref{fig6}), the luminosity (Table~\ref{tbl3}), and the semimajor axes (Fig.~\ref{fig7}).
\\
\item[(2)~] Growth time-scales of the envelope:\\ \\
We have simulated the growth of the protoplanetary envelope after the core growth stops. 
Our simulations demonstrate that the envelope pollution can also hasten gas accretion on to the protoplanet (Fig.\ref{fig8}), which implies that envelope pollution may enable gas giant formation with small cores (Fig.~\ref{fig9} and Table~\ref{tbl4}). 
\end{itemize}
\vspace{1em}

To confirm that our idea really works, further studies will be required: (i) chemical evolution in the polluted envelope coupled with eddy diffusion and inflow of disc gas, (ii) dynamical stability of the polluted envelope, (iii) competitive core growth (planetary accretion) beyond the snow line in a multiple-protoplanet system, (iv) the efficiency in mass deposits of icy planetesimals based on their trajectories and sizes in the envelope, (v) envelope pollution due to erosion of a core itself, and (vi) chemical equilibria of non-ideal gases in the polluted envelope. However, it is quite certain that envelope pollution by planetesimals occurs and affects the formation process of gas giants. We claim that it is necessary to take into account envelope pollution when we discuss gas giant formation based on the-core accretion model.

\section{Acknowledgements \label{sec.7}}

We thank S. Ida for fruitful discussion and several supports. 
We appreciate that J. Ferguson calculated and sent the gas opacity data on his request. 
We also thank T. Guillot and M. Podolak for their critical comments in the early phase of this study and an anonymous referee for the useful comments on this manuscript. Y.H. is supported by Grant-in-Aid for JSPS Fellows (No.21009495)
from the Ministry of Education, Culture, Sports, Science and
Technology (MEXT) of Japan.

\label{lastpage}
\end{document}